# Ultrafast and efficient coherence creation in Λ- like atomic systems driven by nonlinearly chirped few-cycle pulses

Amarendra K. Sarma* and Parvendra Kumar
Department of Physics, Indian Institute of Technology Guwahati, Guwahati-781039, Assam, India.
*Electronic address: aksarma@iitg.ernet.in

We report an ultrafast and efficient way to create the maximum coherence between the two lower states in a Λ-like atomic systems, driven by two nonlinearly chirped few-cycle pulses. The phenomenon of coherent population trapping and electromagnetically induced population transfer are investigated by solving the appropriate density matrix equations without invoking the rotating wave approximation. The robustness of the scheme for maximum coherence against the variation of the laser parameters are tested numerically. We also demonstrate that the proposed scheme could be used to obtain complete population transfer to an initially empty ground state.

Recent progress in ultrafast optics enables generation of few-cycle optical pulses with the durations of only a few periods of the optical radiation [1]. With the advent of these few-cycle pulses research in light-matter interaction is getting a tremendous boost owing to many potential applications [2-3]. In particular, high harmonic generation and attosecond x-ray generation using few-cycle pulses is of extreme technological importance. In this context, the generation of coherence between the ground and an excited state is a very important issue due to the fact that the coherently prepared atomic medium is used for generation of high-order harmonics or multi-photon ionization [4]. On the other hand, coherent population transfer and coherent creation of superposition of quantum states in atoms and molecules have found a number of new applications ranging from coherent population trapping [5-6], control of chemical reactions [7-8], electromagnetically induced transparency (EIT) to quantum information processing and so on[8-18]. Recently, EIT and maximum coherence creation has been reported with a train of ultra-short pulses [9, 10]. Moreover, quantum control of physical and chemical processes and also of attosecond electronic dynamics by use of frequency- and amplitude-chirped few-cycle pulses is also explored by some authors [11, 12]. Motivated by these developments, in this work, we demonstrate an ultrafast, efficient way to create the maximum coherence and almost complete population transfer to the two lower states in a Λ-like three level atomic system driven by two nonlinearly chirped few-cycle laser pulses. In our proposed scheme the maximum coherence and almost complete population transfer could be achieved, just by a judicious choice of the electric field amplitudes and the nonlinear chirp parameters of the given nonlinearly chirped few-cycle pulses. The creation of maximum coherence within few femtoseconds may have potential applications in controlling chemical reactions. It may be noted that, recently, it has been pointed out by many authors that the so-called rotating wave approximation (RWA) do not hold when one deals with few-cycle pulse related phenomena and should work in the non-RWA regime [3,13-15]. In this work we are working in the non-RWA regime and assume that all the atomic relaxation times are considerably longer than the interaction time with the laser pulses. Our analysis is based on the scheme depicted in Fig.1. Here we consider a Λ-like atomic systems interacting with two few-cycle laser pulses. The electric field of the circularly polarized laser interacting between $|3\rangle$ and $|1\rangle$ is given by $\vec{E}_1 = \vec{E}_{10} \exp(-t^2/\tau^2) \cos(\omega_{10} t + \chi_1 t^3)$, where $\vec{E}_{10}, \tau, \omega_{10}$ and $\chi_1$ are respectively, the amplitude, the temporal width, the carrier frequency and the chirp rate of the pulse. Exactly analogous expression for the linearly polarized laser pulse interacting between $|3\rangle$ and $|2\rangle$ is given by $\vec{E}_2 = \vec{E}_{20} \exp(-t^2/\tau^2) \cos(\omega_{20} t + \chi_2 t^3)$. In the proposed scheme, atomic states, $|1\rangle$ $(3\,^2S_{1/2}(F=1))$, $|2\rangle$ $(3\,^2S_{1/2}(F=2))$ and $|3\rangle$ $(3\,^2P_{1/2}(F=2))$ represent the hyperfine quantum states of sodium atom. In the given scheme, we assume that only $|3\rangle \to |1\rangle$ and $|3\rangle \to |2\rangle$ transitions are dipole allowed while $|2\rangle \to |1\rangle$ transitions are forbidden.

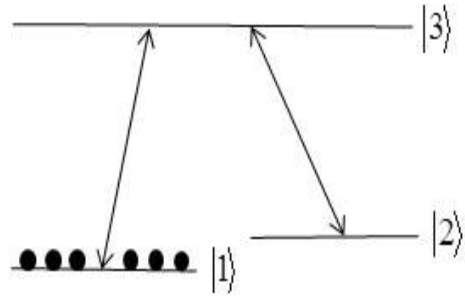

Fig. 1 Schematic of $\Lambda$-like system with two acting nonlinearly chirped few-cycle laser pulse

The Hamiltonian of the system is given by $\hat{H} = \hat{H}_0 + \hat{H}_{int}$ where $\hat{H}_0 = \hbar\omega_1 |1\rangle\langle 1| + \hbar\omega_2 |2\rangle\langle 2| + \hbar\omega_3 |3\rangle\langle 3|$ and
$\hat{H}_{int} = -\vec{\mu}.\vec{E} = -\hbar\Omega_{31}(t)|3\rangle\langle 1| - \hbar\Omega_{32}(t)|3\rangle\langle 2| + $ h.c.
Here $\Omega_{31}(t) = \mu_{31}\vec{E}_1(t)/\hbar$ and $\Omega_{32}(t) = \mu_{32}\vec{E}_2(t)/\hbar$ are the time dependent Rabi frequency for the transition with electric dipole moment $\mu_{31}$ and $\mu_{32}$ respectively. The Bloch equations, without invoking the so called rotating wave approximation, describing the temporal evolution of the density matrix elements are:

$$\dot{\rho}_{31} = -i\omega_{31}\rho_{31} + i\Omega_{32}(t)\rho_{21} - i\Omega_{31}(t)(\rho_{33} - \rho_{11})$$
$$\dot{\rho}_{32} = -i\omega_{32}\rho_{32} + i\Omega_{31}(t)\rho_{12} - i\Omega_{32}(t)(\rho_{33} - \rho_{11})$$
$$\dot{\rho}_{21} = -i\omega_{21}\rho_{21} + i\Omega_{32}(t)\rho_{31} - i\Omega_{31}(t)\rho_{23}$$
$$\dot{\rho}_{11} = i\Omega_{31}(t)(\rho_{31} - \rho_{13})$$
$$\dot{\rho}_{22} = i\Omega_{32}(t)(\rho_{32} - \rho_{23})$$
$$\dot{\rho}_{33} = i\Omega_{31}(t)(\rho_{13} - \rho_{31}) + i\Omega_{32}(t)(\rho_{23} - \rho_{32})$$

(1)

Here $\omega_{ij} = \omega_i - \omega_j$. It may be noted that $\rho_{ij} = \rho_{ji}^*$. We solve Eq. (1) numerically using a standard fourth-order Runge-Kutta method. We assume that initially all the atoms are in the ground state $|1\rangle$. We use the following typical parameters: $\omega_{31} = \omega_{10} = 3.18$ rad/fs, $\omega_{21} = 0.00001$ rad/fs, $\omega_{32} = \omega_{20} = 3.1799$ rad/fs, $\tau = 4.49$ fs and $\mu_{31} \approx \mu_{13} = 2.49\,ea_0$, where e and $a_0$ are the electron charge and the Bohr radius respectively. It may be noted that the chosen laser field parameters in this work are very close to that of a sodium (Na) atom. In Fig. 2(a) we depict the coherence, $\rho_{12}$, between the two lower states during the interaction of the two laser pulses and populations, $\rho_{11}, \rho_{22}$ and $\rho_{33}$ with $\Omega_{10} = 1$ rad/fs, $\Omega_{20} = 2.4$ rad/fs and $\chi_1 = \chi_2 = 0.397\,\text{fs}^{-3}$. On the other hand, Fig.2 (b) depicts the dark and bright states populations pertaining to the lower states. It may be noted that, here $\Omega_{10}$ and $\Omega_{20}$ are the peak Rabi frequencies associated with $\vec{E}_1$ and $\vec{E}_2$ respectively.

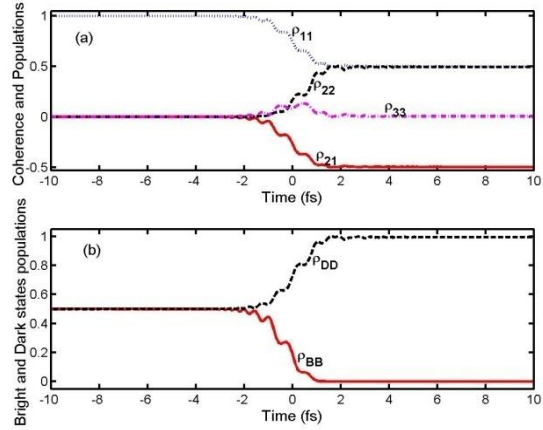

Fig. 2 (Color online) (a) Coherence between the two lower states $\rho_{21}$ and populations $\rho_{11}, \rho_{22}$ and $\rho_{33}$ with $\Omega_{10} = 1$ rad/fs, $\Omega_{20} = 2.4$ rad/fs and $\chi_1 = \chi_2 = 0.397$ fs$^{-3}$. (b) Bright and dark state populations $\rho_{BB}$ and $\rho_{DD}$ corresponding to (a).

It can be clearly seen that maximum coherence of 0.5 could be obtained between the two lower states as a result of the interaction for the parameters chosen. Also the population of the excited state becomes zero while the two lower states become equally populated. Keeping these results in view, it may be plausible to conclude that we have established electromagnetically induced transparency in the excitation of a $\Lambda$-like system by two nonlinearly chirped pulses. In fact, this is confirmed by the results depicted in Fig. 2(b). Here we observe that, though both the dressed states started out with equal population, as a result of interaction the population in the bright state is transferred nearly completely to the dark state where it remains trapped. It is worthwhile to mention that such coherent accumulation of excitation and EIT have been numerically observed with an ultrashort pulse train [9]. On the other hand, for the first time, to the best of our knowledge, we are reporting such excitation of $\Lambda$-like atomic systems by a couple of nonlinearly chirped single few-cycle pulses.

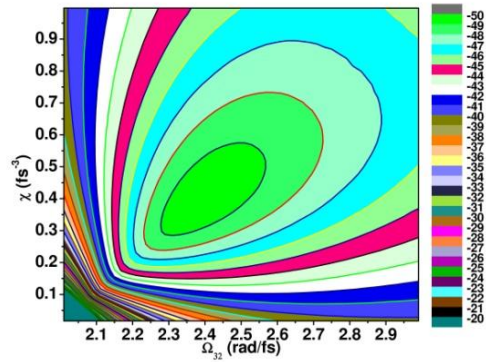

Fig. 3 (Color online) Contour plot of coherence as a function of nonlinear chirp parameter ($\chi_1 = \chi_2 = \chi$) and Rabi frequency, $\Omega_{32}$ at a constant $\Omega_{31}$.

In order to verify the robustness of our scheme, with respect to the nonlinear chirp parameter and peak Rabi frequency between corresponding to $|3\rangle \rightarrow |2\rangle$ transition, in Fig. 3, we plot the contour map for coherence $\rho_{21}$ as a function of $\Omega_{32}$ and $\chi$. One can observe that nearly maximum coherence of 0.5 or 50% is obtained for a fairly wide variations in both the Rabi frequency and the nonlinear chirp parameter. Our analysis shows that the coherence is fairly robust against the variations in the laser pulse duration and $\Omega_{31}$ also. A near complete population transfer could be achieved by a judicious choice of the laser parameters, as may be observed from Fig. 4 where we depict the populations $\rho_{11}, \rho_{22}$ and $\rho_{33}$ with $\Omega_{10} = 1.67$ rad/fs, $\Omega_{20} = 2.5$ rad/fs, $\chi_1 = 0.6$ fs$^{-3}$ and $\chi_2 = 0.4$ fs$^{-3}$. This is a remarkable result in the sense that, with the same scheme, subject to the appropriate choice of the laser parameters one may obtain either complete population transfer or an EIT. In fact we have found that this scheme is equally applicable to other alkali atomic systems like Li, Rb and Cs also.

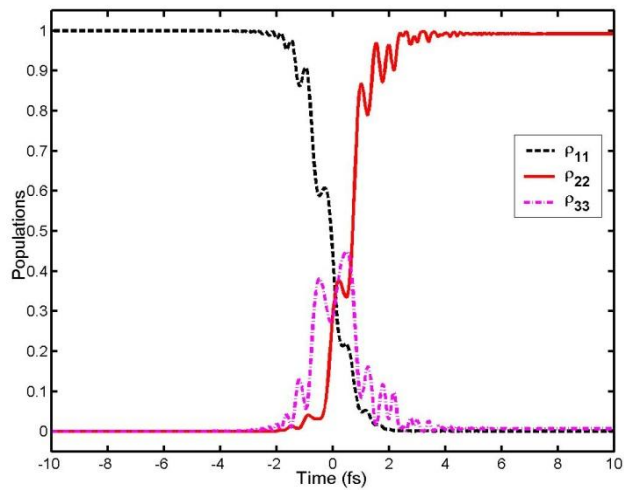

Fig. 4 (Color online) Populations $\rho_{11}, \rho_{22}$ and $\rho_{33}$ with $\Omega_{10} = 1.67$ rad/fs, $\Omega_{20} = 2.5$ rad/fs, $\chi_1 = 0.6$ fs$^{-3}$ and $\chi_2 = 0.4$ fs$^{-3}$.

      To conclude, we have demonstrated the ultrafast, efficient and robust, creation of coherence between the ground states of Na-like atoms. The phenomenon of coherent population trapping is confirmed by the dynamics of the bright and the dark state population. This leads to the phenomenon of electromagnetically induced transparency of a weak resonant pulse between quantum states $|1\rangle$ and $|3\rangle$. The induced coherence is fairly robust against the variations in nonlinear chirp parameters, Rabi frequencies and in pulse durations. We have also demonstrated almost complete population transfer to the initially empty ground state by properly choosing the laser field parameters. We hope that our proposed scheme would find potential applications in many areas such as, the coherent control of chemical reactions and quantum information processing.